\documentclass[showpacs,amsmath,amssymb,aps,showkeys,floatfix,prd,a4paper]{revtex4}

\usepackage[dvips]{graphicx}
\usepackage{dcolumn}
\usepackage{bm}
\usepackage{epsfig}
\usepackage{amsfonts}
\usepackage{amssymb,amscd}

\def\lsim{\raise0.3ex\hbox{$<$\kern-0.75em\raise-1.1ex\hbox{$\sim$}}}
\def\gsim{\raise0.3ex\hbox{$>$\kern-0.75em\raise-1.1ex\hbox{$\sim$}}}

\def\beq{\begin{equation}}
\def\eeq{\end{equation}}
\def\bea{\begin{eqnarray}}
\def\eea{\end{eqnarray}}
\def\bq{\begin{quote}}
\def\eq{\end{quote}}

\def\gappeq{\mathrel{\rlap {\raise.5ex\hbox{$>$}}
{\lower.5ex\hbox{$\sim$}}}}

\def\lappeq{\mathrel{\rlap{\raise.5ex\hbox{$<$}}
{\lower.5ex\hbox{$\sim$}}}}

\def\Toprel#1\over#2{\mathrel{\mathop{#2}\limits^{#1}}}

\begin{document}


\title{$\eta_c$ photoproduction at LHC energies}

\author{V.~P. Gon\c{c}alves and B. D. Moreira}
\affiliation{High and Medium Energy Group, \\
Instituto de F\'{\i}sica e Matem\'atica, Universidade Federal de Pelotas\\
Caixa Postal 354, CEP 96010-900, Pelotas, RS, Brazil}

\date{\today}

\begin{abstract}

In this contribution, we study the inclusive and diffractive $\eta_{c}$ photoproduction in $pp$ and $pPb$ ultra-peripheral collisions 
(UPC's) at the LHC Run 2 energies. The quarkonium production is studied using nonrelativistic quantum chromodynamics (NRQCD) formalism.
We present predictions for rapidity and transverse momentum distributions for the $\eta_c$ 
photoproduction and present our estimate for the total cross sections at the Run 2 energies.

\end{abstract}

\maketitle

\section{Introduction}
\label{intro}

\begin{figure}[t]
\includegraphics[scale=0.45]{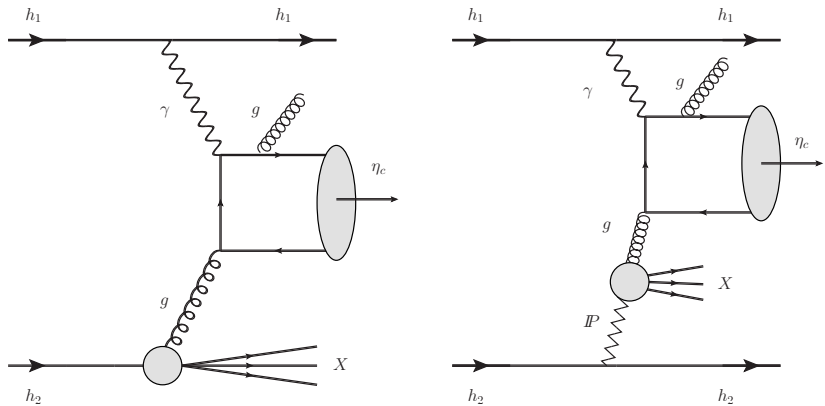}
\caption{Schematic view of   diagrams for the inclusive (left) and diffractive (right) $\eta_c$ production in hadronic collisions considering
photon - hadron interactions.}
\label{eta_c_fig}
\end{figure}

In a hadron-hadron UPC, it is well known that the hadrons can act as sources of almost real photons allowing 
the study of photon-photon and photon-hadron interactions (the photon-induced processes) \cite{upc}. Our goal 
in this contribution is the study of the $\eta_{c}$ production 
in the photon-gluon and the photon-Pomeron 
subprocesses, as shown in Fig.\ref{eta_c_fig}, using the nonrelativistic QCD formalism (NRQCD). In this formalism, the photon-hadron process can be factorized in terms of the short
distance coefficients for the photon-gluon subprocess (perturbatively calculated) and long distance matrix elements, related to the formation of 
the quarkonium at the final state, which are extracted from global analysis of the quarkonium data \cite{haoprl99,elements} . As the photoproduction of $\eta_{c}$ is a purely 
color octet contribution \cite{haoprl99},  this process is a probe of the color  octet mechanism. In our analysis, we present our predictions for the rapidity and 
$p_{T}$ distributions and total cross sections for the $\eta_{c}$ photoproduction. A comparison with the predictions for the $\eta_{c}$ production in
photon-photon and exclusive photon-hadron interactions also is presented. 

This contribution is organized as follows. In the next section we present the formalism to study the $\eta_{c}$ photoproduction at hadron-hadron 
collisions, presenting a brief summary on the equivalent photon approximation and NRQCD. In Section \ref{res}, we summarize our results for 
rapidity and $p_{t}$ distributions and total cross sections. A more detailed discussion is presented in Ref.\cite{Goncalves:2018yxc}. Lastly, 
in Section \ref{conclusions}, we present our conclusions.

\section{$\eta_{c}$ production at photo-induced processes}
\label{formalis}

Let us consider an ultraperipheral collision (impact parameter $ > R_{h_{1}}+R_{h_{2}} $) between two fast hadrons. In this regime we can 
factorize the cross section for th $\eta_{c}$ production in $hh$ collisions in terms of the equivalent flux of photons of the hadron projectile and the 
photon-hadron cross section. 
The rapidity distribution for quarkonium photoproduction in hadron-hadron collisions is given by \cite{Goncalves:2018yxc}
\begin{eqnarray}
\frac{d\sigma_{hh}}{dY}(Y) = n_{h_{1}}(Y) \sigma_{\gamma h_{2}}(Y) + \left(   1   \longleftrightarrow 2 \,; Y \rightarrow -Y \right)   
\end{eqnarray}
where the rapidity variable is related with the photon energy by $Y = \ln(2 \omega / M_{\eta_{c}})$ and
$n(Y)$ is the photon spectrum associated to the proton or nucleus which is dependent of the choice of the form factor. For protons as source of 
photons we use the dipole form factor, which leads to \cite{Dress}
\begin{eqnarray}
n_{p} (\omega) =  \frac{\alpha_{\mathrm{em}}}{2 \pi\, } \left[ 1 + \left(1 -
\frac{2\,\omega}{\sqrt{s_{NN}}}\right)^2 \right] 
\left( \ln{\Omega} - \frac{11}{6} + \frac{3}{\Omega}  - \frac{3}{2 \,\Omega^2} + \frac{1}{3 \,\Omega^3} \right) \,,
\label{eq:photon_spectrum}
\end{eqnarray}
where $\Omega = 1 + [\,(0.71 \,\mathrm{GeV}^2)/Q_{\mathrm{min}}^2\,]$ and 
$Q_{\mathrm{min}}^2= \omega^2/[\,\gamma_L^2 \,(1-2\,\omega /\sqrt{s})\,] \approx (\omega/
\gamma_L)^2$, where $\gamma_L$ is the Lorentz boost  of a single beam. For $Pb$  we have used the hard sphere form factor, which gives 
the following photon flux \cite{upc}
\begin{eqnarray}
n_A(\omega) = \frac{2\,Z^2\alpha_{em}}{\pi\,}\, \left[\bar{\eta}\,K_0\,(\bar{\eta})\, K_1\,(\bar{\eta}) - \frac{\bar{\eta}^2}{2}\,{\cal{U}}(\bar{\eta}) \right]\,
\label{fluxint}
\end{eqnarray}
where   $K_0(\eta)$ and  $K_1(\eta)$ are the
modified Bessel functions, $\bar{\eta}=\omega\,(R_{h_1}+R_{h_2})/\gamma_L$ and  ${\cal{U}}(\bar{\eta}) = K_1^2\,(\bar{\eta})-  K_0^2\,(\bar{\eta})$.

In order to study the inelastic $\eta_{c}$ photoproduction in the process $\gamma + p \rightarrow \eta_{c} + X$, we will use the NRQCD 
formalism, which allows us to factorize the subprocess $\gamma + g \rightarrow \eta_{c} + g$ in two different steps: (i) the heavy pair production 
(perturbative), and (ii) the $\eta_{c}$ formation, which is a nonperturbative process (see Fig. \ref{eta_c_fig}).
Firstly, for the inclusive process 
$\gamma + p \rightarrow \eta_{c} + X$ the cross section is given by \cite{Goncalves:2018yxc, haoprl99}
 \begin{eqnarray}
\sigma (\gamma + h \rightarrow \eta_c + X ) = \int dz dp_{T}^2 \frac{xg(x,Q^2)}{z(1-z)} 
 \frac{d\sigma}{d \hat{t} }(\gamma + g \rightarrow \eta_c + g ) \label{sigmagamp}
\end{eqnarray}
where $z \equiv (p_{\eta_c}.p)/(p_{\gamma}.p)$ (fraction of the photon energy carried away by the ${\eta_c}$ in the hadron rest frame), with 
$p_{\eta_c}$, $p$ and $p_{\gamma}$ being the four momentum of the ${\eta_c}$, hadron and photon, 
respectively. $p_{T}$ is 
the magnitude of the $\eta_c$ transverse momentum and $g(x,Q^2)$ is the standard gluon distribution function, which will be modelled using the
CTEQ6LO parametrization \cite{cteq6} assuming that 
$Q^2 = 4 m_c^2$. For the diffractive case we have used the diffractive gluon PDF derived by the H1 collaboration in  Ref.\cite{H1diff}. The integration 
limits are taken so that $z < 1$ and $p_{T}^{2} \ge 1$ GeV$^{2}$.
The associated partonic differential cross section $d\sigma/d \hat{t}$  is given by (see Ref.\cite{haoprl99}) 
\begin{eqnarray}
 \frac{d\sigma}{d\hat{t}} = \frac{1}{16\pi \hat{s}^{2}} 
 F (^{2S+1}L_{J}^{[8]})   
\times      \langle {\cal O}(^{2S+1}L_{J}^{[8]})  \rangle \,\,,
\label{dsigdt_gamma_gluon} 
 \end{eqnarray}
with the short distance coefficients $F$ of the subprocesses being given by \cite{haoprl99}
\begin{eqnarray}
 F(^{3}S_{1}^{[8]}) &=& 20(4\pi)^{3} \alpha \alpha_{S}^{2} e_{c}^{2} M  
 \frac{{\cal P}^{2} - M^{2}\hat{s}\hat{t}\hat{u}}{9{\cal Q}^{2}}         \\
 F(^{1}S_{0}^{[8]}) &=& 3(4\pi)^{3} \alpha \alpha_{S}^{2} e_{c}^{2} \hat{s} \hat{u}  
 \frac{M^{8} + \hat{s}^{4} + \hat{t}^{4} + \hat{u}^{4}} {M \hat{t} {\cal Q}^{2}}         \\
 F(^{1}P_{1}^{[8]}) &=& \frac{80 (4\pi)^{3} \alpha \alpha_{S}^{2} e_{c}^{2} }{9M {\cal Q}^{3}}
 \left[  
M^{2}{\cal Q} \left( M^{6} + 5 \hat{s}\hat{t}\hat{u} - {\cal Q}     \right) \right.  \nonumber \\
 &-&\left.    2\hat{s}\hat{t}\hat{u}\left( {\cal P}^{2} + 2M^{8} - M^{2} \hat{s}\hat{t}\hat{u}          \right)
 \right] \,\,\, ,
\end{eqnarray}
and
\begin{eqnarray}
 {\cal P} &=& \hat{s} \hat{t} + \hat{t} \hat{u} + \hat{s} \hat{u}         \,\, ,  \\
 {\cal Q} &=& (\hat{s} + \hat{t}) (\hat{s} + \hat{u}) (\hat{t} + \hat{u}).
\end{eqnarray}
Moreover, $\langle {\cal O}(^{2S+1}L_{J}^{[8]})  \rangle$ are the long distance matrix elements, 
obtained from experimental data of the quarkonium production. Here we have used the updated values from Ref.\cite{elements}.


\section{Results}
\label{res}

\begin{figure}
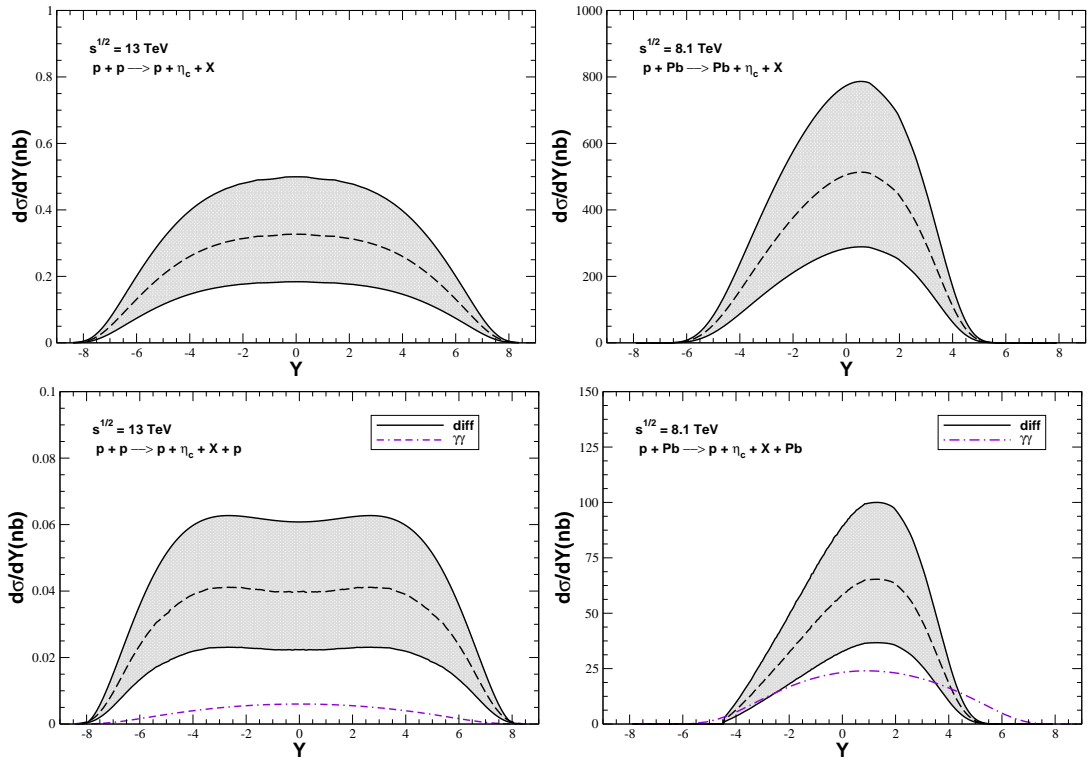

\begin{tabular}{cc}
\includegraphics[scale=0.3]{BANDAS_pp_TOT_etaC_dsigdy_17.eps} & 
\includegraphics[scale=0.3]{BANDAS_pA_TOT_etaC_dsigdy_17.eps}  \\
\includegraphics[scale=0.3]{BANDAS_diff_pp_TOT_etaC_dsigdy_17.eps} & 
\includegraphics[scale=0.3]{BANDAS_diff_pA_TOT_etaC_dsigdy_17.eps}
\end{tabular}
\caption{Rapidity distribution for the $\eta_{c}$ photoproduction.  
Upper row: inclusive photoproduction in $pp$ collisions (left panel) and $pA$ collisions (right panel). 
Lower row: diffractive photoproduction in $pp$ collisions (left panel) and $pA$ collisions (right panel).}
\label{dsigdy}
\end{figure}

In the Figure \ref{dsigdy} we present our predictions for the rapidity distribution ($pp$ at $\sqrt{s}=13$ TeV and $pA$ at $\sqrt{s}=8.1$ TeV). As 
expected, the rapidity distributions 
in $pp$ collisions are symmetric in $Y=0$, while in the $pA$ case are asymmetric because of the dominance of the $\gamma p$ interactions when 
the nucleus is 
acting as a source of photons. The bands presented 
are due to the uncertainty of the long distance matrix elements (Ref.\cite{elements}). Moreover, the inclusive production 
is about 10 times larger than the diffractive production. For the diffractive cases we present also the $\gamma \gamma $ $\eta_{c}$ 
photoproduction, from Ref.\cite{Goncalves:2018yxc}. 

\begin{figure}[t]
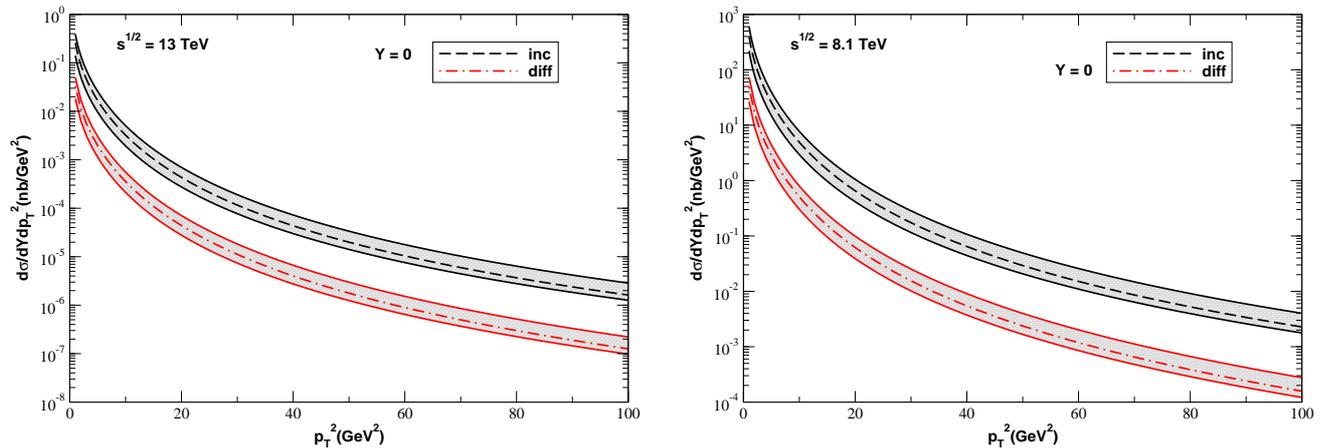

\begin{tabular}{ccc}
\includegraphics[scale=0.35]{BANDAS_dsigdydpt2_TOT_pp_etaC_17.eps} & \,\,\,\,\, & 
\includegraphics[scale=0.35]{BANDAS_dsigdydpt2_TOT_pA_etaC_17.eps}
\end{tabular}
\caption{Transverse momentum distributions for the inclusive and diffractive $\eta_c$  photoproduction at central rapidities ($Y = 0$) in
$pp$ collisions at $\sqrt{s} = 13$ TeV (left panel) and  $pPb$  collisions at $\sqrt{s} = 8.1$ TeV (right panel).}
\label{dist_pt}
\end{figure}

In Fig. \ref{dist_pt} we present the transverse momentum distributions 
for $\eta_{c}$ photoproduction at $Y = 0$ and $\sqrt{s}=13$ TeV. The dashed lines represent 
the inclusive case and the dot-dashed lines the diffractive case. As in Figure 2, we 
can observe that the inclusive case is about 10 times larger that the diffractive 
case. Moreover, we can observe that the behavior of distribution is $1/p_{t}^{n}$, which is a typical DGLAP behavior.

In the Table 1 we present our predictions for the total cross sections for the $\eta_{c}$ photoproduction. For comparison, 
we present the predictions for the $\eta_{c}$ photoproduction due to $\gamma \gamma$ and $\gamma I\!\!\! O$ mechanisms, from 
Refs.\cite{Goncalves:2018yxc,vic_odderon}. 
As can be observed, the $\gamma p$ inclusive mechanism is the dominant mechanism for the $\eta_{c}$ photoproduction.
Here it is important to emphasize 
that the $\gamma I\!\!P$ diffractive, the $\gamma \gamma$ and the $\gamma I\!\!\!O$ processes are characterized by two rapidity gaps at the final state. 
Furthermore, for the $\gamma I\!\!\!O$ mechanism we have the smaller cross sections.


\hspace{2cm}

\begin{table}[h!]
\centering
\begin{tabular}{||c||c||c||c||c||}\hline

                              & \,  $\gamma p$  inclusive   \,                    &  \,    $\gamma I\!\! P$   diffractive     \,             &   \, $\gamma \gamma$ exclusive  \,  &  \,  $\gamma I\!\!\! O $ exclusive     \,        \\ \hline
$pp$ ($\sqrt{s} = $ 13 TeV)   &    3.492 nb            &     0.501 nb               &    0.059 nb  &  0.013 nb \\ \hline
$pPb$ ($\sqrt{s} = $ 8.1 TeV) &    3.194  $\mu$b     &  0.351 $\mu$b         &   0.182 $\mu$b     &  0.032 $\mu$b                   \\ \hline
\end{tabular} 
\caption{Total cross section for the $\eta_{c}$ photoproduction. Here we also present the 
$\eta_{c}$ production due to $\gamma \gamma $ and $\gamma I\!\!\! O$ mechanisms from Refs.\cite{Goncalves:2018yxc,vic_odderon}}
\end{table}


\section{Conclusions}
\label{conclusions}
The photo-induced processes at LHC have been a powerful tool to study the hadronic structure and the QCD
dynamics in different processes. In this contribution, we have studied the $\eta_{c}$ photoproduction in the NRQCD formalism.
Following the Ref.\cite{haoprl99} , this process is a 
pure color octet and, therefore, it is a probe of the color octet mechanism. 
Moreover, it is important to note that the diffractive, $\gamma \gamma$ and $\gamma-$Odderon mechanisms are characterized by two rapidity gaps at the 
final state. In 
particular, the detection 
of the exclusive ($\gamma - $Odderon) mechanism would be a direct probe of the perturbative Odderon. Since the diffractive
and $\gamma \gamma$ processes are backgrounds for the $\gamma - $Odderon process, the knowledge of their magnitudes can be 
useful  for experimental analysis. 
Finally, the magnitude of the cross sections presented here, show that this study is feasible at LHC energies.

\section*{Acknowledgements}
This work was  partially financed by the Brazilian funding
agencies CNPq, CAPES,  FAPERGS and  INCT-FNA (process number 
464898/2014-5).



\end{document}